\documentclass{aa}

\usepackage{natbib}
\bibpunct{(}{)}{;}{a}{}{,} 
\usepackage{graphicx}
\usepackage{txfonts}

\newcommand{\powerlaw}{{\fontfamily{ptm}\fontseries{m}\fontshape{sc}\selectfont{powerlaw}}}

\newcommand{\zphabs}{{\fontfamily{ptm}\fontseries{m}\fontshape{sc}\selectfont{zphabs}}}
\newcommand{\tbabs}{{\fontfamily{ptm}\fontseries{m}\fontshape{sc}\selectfont{tbabs}}}

\newcommand{\xstar}{{\fontfamily{ptm}\fontseries{m}\fontshape{sc}\selectfont{xstar}}}

\newcommand{\zgauss}{{\fontfamily{ptm}\fontseries{m}\fontshape{sc}\selectfont{zgauss}}}
\newcommand{\reflion}{{\fontfamily{ptm}\fontseries{m}\fontshape{sc}\selectfont{reflion}}}

\newcommand{\kyconv}{{\fontfamily{ptm}\fontseries{m}\fontshape{sc}\selectfont{kyconv}}}

\begin{document}
\title{Warm absorber and truncated accretion disc in IRAS 05078+1626}
\author{J.~Svoboda,$\!^{1,2}$ M.~Guainazzi,$\!^{3}$ \and V.~Karas$^{1}$}
\institute{$^{1}$ Astronomical Institute, Academy of Sciences, Bo\v{c}n\'{\i}~II~1401, CZ-14131~Prague, Czech~Republic \\ 
$^{2}$ Faculty of Mathematics and Physics, Charles University, Ke Karlovu~3, CZ-12161~Prague, Czech~Republic \\
$^{3}$ European Space Astronomy Centre of ESA, PO Box 78, Villanueva de la Ca\~{n}ada, 28691 Madrid, Spain}

\authorrunning{J.~Svoboda et al.}
\titlerunning{Warm absorber and truncated accretion disc in IRAS 05078+1626}
\date{Received 13 November 2009 / Accepted 24 December 2009}
\offprints{J.~Svoboda, email: svoboda@astro.cas.cz}
\abstract{X-ray observations of unabsorbed active galactic nuclei provide an opportunity 
to explore the innermost regions of supermassive black hole accretion discs.}{Our
goal in this paper is to investigate the central environment of
a Seyfert 1.5 galaxy IRAS 05078+1626.} 
{We studied the time-averaged spectrum obtained with the EPIC and RGS
instruments onboard XMM-Newton.}
{A power law continuum (photon index $\Gamma\simeq1.75$) 
dominates the 2--10~keV energy range. A narrow iron K\,$\alpha$ spectral line is clearly seen,
presumably originating in a distant torus, but no broad relativistic component was 
detected. However, the power law and the iron K\,$\alpha$ line alone do not provide
a satisfactory fit in the soft X-ray band whose spectrum can be explained by the combination
of three components: a) \textit{a cold photoelectric absorber} 
with column density $N_{H} \approx 10^{21}$cm$^{-2}$.
This gas could be located either in outer parts of the accretion disc, at the rim 
of the torus or farther out in the host galaxy; 
b) \textit{a warm absorber} with high ionization parameter ($\log \xi \approx 2.2$)
and column density ($N_{H} \approx 10^{24}$cm$^{-2}$); c) \textit{an ionised reflection}
where the reflecting gas could be either in the inner wall of a warm absorber cone
or in an ionised accretion disc.} 
{The first X-ray spectroscopic measurement 
of IRAS05078+1626 unveils some of the standard ingredients in Seyfert galaxies, 
such as a power law primary continuum, modified by reflection from the accretion disc 
and by the effect of complex, multi-phase obscuration. However, data constrains the 
accretion disc, if present, not to extend closer than to 
60 gravitational radii from the black hole.}
\maketitle

\section{Introduction}

IRAS 05078+1626 is a nearby Seyfert 1.5 galaxy. 
Before its identification as an infrared source
it was also known as CSV 6150 (Catalogue of Suspected Variables). 
Its position on the sky is $l=186.1$ and $b=-13.5$ in the Galactic coordinates. 
The cosmological redshift of this galaxy is $z\approx 0.018$ \citep{takata94}. 
It had never been spectroscopically examined in X-ray prior to the
observation discussed in this paper.
However, it was detected in a number of X-ray surveys, such as the
all-sky monitoring of the INTEGRAL IBIS/ISGRI instrument \citep{sazonov07}, 
the SWIFT BAT instrument \citep{ajello08,2008ApJ...681..113T}, and the RXTE Slew Survey (XSS) \citep{2004A&A...418..927R}. 

The X-ray spectroscopic properties of intermediate Seyferts are rather elusive:
both obscured Type 1 and unobscured Type 2 active galactic nuclei (AGN) have been reported
\citep[e.g.,][]{2006A&A...446..459C,2008MNRAS.390.1241B,2009ApJ...695..781B}.
It has been suggested that intermediate Seyfert galaxies are seen at intermediate 
inclination angles between pure ``face-on'' Seyfert~1s and pure ``edge-on'' Seyfert~2s,
which follows directly from the orientation-based AGN unification scenarios 
\citep{1985ApJ...297..621A,1993ARA&A..31..473A,1995PASP..107..803U}.
For this reason, X-ray spectroscopy of type 1.5 Seyferts may provide clues
to the nature and geometrical distributions of optically thick gas surrounding
the active nucleus, the latter being the fundamental ingredient behind 
the unification scenarios.

IRAS~05078+1626 is included in the FERO project 
(``Finding Extreme Relativistic Objects''; \citeauthor{2008MmSAI..79..259L}
\citeyear{2008MmSAI..79..259L}) 
with the aim of establishing the fraction of a relativistically broadened 
K$\alpha$ iron lines in the spectrum of a complete flux-limited sample (2--10\,keV flux $>$ 1 mCrab).

This paper is organised as follows. In section~\ref{sec:observations} we describe
new observation by XMM-Newton and the corresponding data reduction.
In sections~\ref{time}~-~\ref{res}, we present the results 
derived from the X-ray spectra in $\sim1$--$10$ keV energy range.
The results are discussed in section~\ref{discussion} and
the conclusions are summarised in section~\ref{conclusions}.

\section{Results from XMM-Newton observation in 2007}
\subsection{Observations and data reduction}
\label{sec:observations}
The XMM-Newton observation of IRAS 05078+1626 was performed between 2007 August 21 
UT 22:24:49 and 22 UT 15:35:43 (Obs. \#0502090501).
The EPN and both MOS cameras (\citet{pn} for PN; \citet{mos} for MOS) were operating 
in the small window mode. The RGS cameras \citep{rgs} were operating in spectroscopic mode.
The spectra were reduced with the SAS software version 9.0.0 \citep{2004ASPC..314..759G}. 
Intervals of high particle background were removed by applying count rate thresholds
on the field-of-view (EPIC, single events) and CCD\,\#9 (RGS) light curves of 0.35\,cts/s for the PN, 
0.5\,cts/s for the MOS and 0.15\,cts/s for the RGS. The exposure time after data
screening is $\approx 56$\,ks for MOS, $\approx 40$\,ks for PN and $\approx 58$\,ks for RGS, respectively.
The patterns 0--12 were used for both MOS cameras, and patterns 0--4 (i.e.\ single and double events) for the PN camera. 
The source spectra were extracted from a circle of 40 arcsec in radius defined around the centroid position 
with the background taken from an offset position close to the source. 
The two MOS spectra and the related response files were joined 
into a single spectrum and response matrix.
Finally, the PN and MOS spectra were rebinned in order to have 
at least 25 counts per bin and to oversample the energy resolution 
of the instrument maximally by a factor of three, while the RGS spectra
were left unbinned. Consequently, different statistics were used in fitting
the spectra -- the traditional $\chi^2$ statistics to fit the PN and MOS spectra
and the C-statistics \citep{1976A&A....52..307C} for all fits including RGS data.
For the spectral analysis, we used XSPEC \citep{1996ASPC..101...17A} 
version 12.5, which is part of the HEASOFT software package version~6.6.\footnote{http://heasarc.gsfc.nasa.gov}

\subsection{Timing properties}
\label{time}

\begin{figure}[thb]
\begin{tabular}{c}
\includegraphics[angle=0,width=0.48\textwidth]{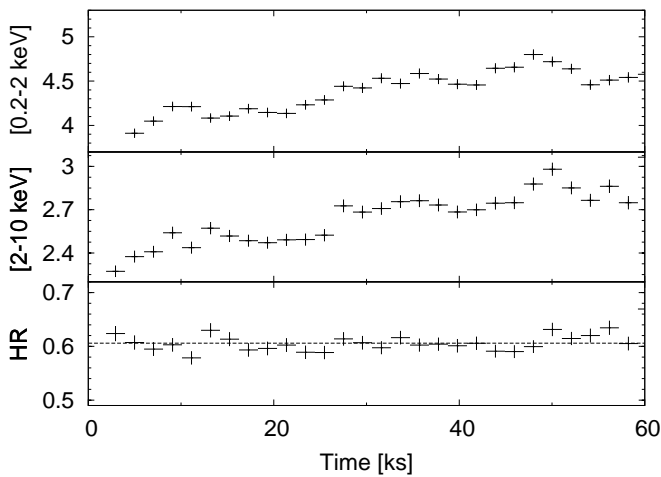} 
\end{tabular}
\caption{EPIC-PN light curves in the 0.2--2\,keV band (upper panel) and 2--10\,keV band (middle). 
The hardness ratio HR is defined as the ratio of the counts at 2--10\,keV to the counts at 0.2--2\,keV
and presented as a function of time in the lower panel. The bin time is as 2048\,s.}
\label{lc}
\end{figure}

The PN light curve of the source is shown in Fig.~\ref{lc}. We have divided the energy range
into two bands and checked the light curve behaviour in each of them, 
as well as a hardness ratio, which
we defined as the ratio of the counts at 2--10\,keV to the counts at 0.2--2\,keV.
The energy ranges were chosen for sampling different spectral components,
as indicated by the energy where the continuum starts deviating from a power law model
that describes the hard X-ray spectrum (see Sect.~\ref{msp}).
The hardness ratio stays almost constant during the observation, 
suggesting that no significant spectral variations occur, 
although the source flux increased by around 20\%. 

\begin{figure}[tb!]
\includegraphics[angle=0,width=0.48\textwidth]{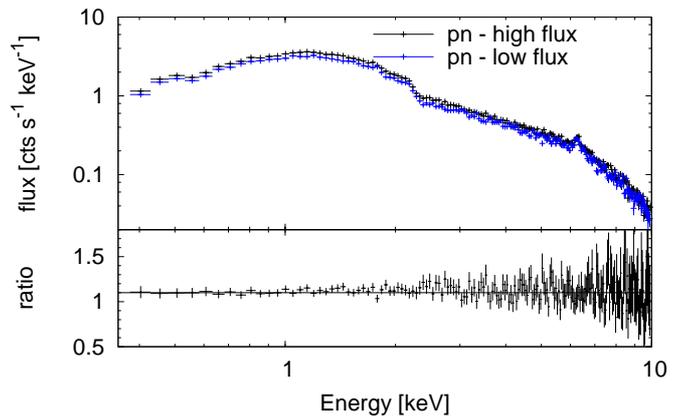} 
\caption{PN spectrum extracted from the first half of the observation with
the lower source flux (blue) and from the second half of the observation with
the higher source flux (black). The ratio of the two spectra is presented
in the lower panel.}
\label{pndata}
\end{figure}

To confirm this conclusion also for narrow spectral features,
such as the iron emission line, we compared the PN spectra extracted during the
first and the second halves of the observation (see Fig.~\ref{pndata}).
The spectra correspond to the lower/higher source flux 
because the flux is increasing nearly monotonically during the observation.
We calculated the ratio values of the two data sets and fit them with a simple
function $f(E)=a \,E+b$ using the least square method. The fitting results are
$a=-0.004 \pm 0.003$ and $b=1.12 \pm 0.01$ with the sum of the residuals $\chi^{2}=111$ 
for 190 degrees of freedom. When we set $a=0$, the fitting results
are comparably good with $b=1.11 \pm 0.01$ and $\chi^{2}=113$. 
The ratio of the spectra is plotted in the lower panel of Fig.~\ref{pndata}. 
Because no significant spectral differences are evident,
we analyse the time-averaged spectra hereafter.

\begin{figure}[tb!]
\includegraphics[width=0.48\textwidth]{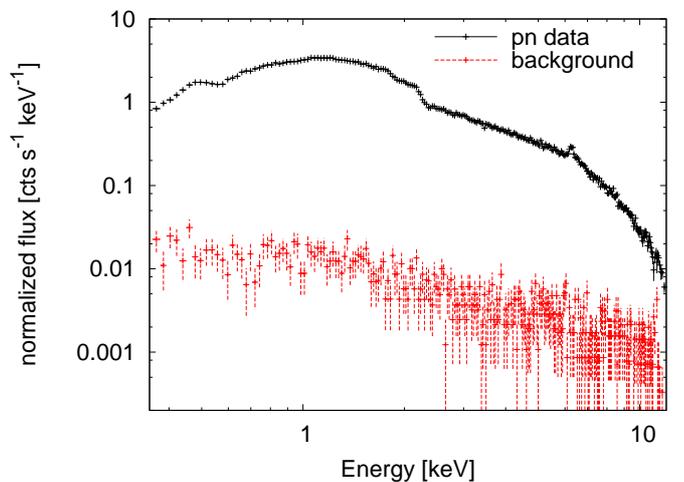}
\caption{Total PN spectrum with the background level showing 
that the signal-to-noise ratio is very good up to high energies.}
\label{pndata_back}
\end{figure}

\begin{figure}[thb]
\begin{tabular}{c}
\includegraphics[angle=0,width=0.48\textwidth]{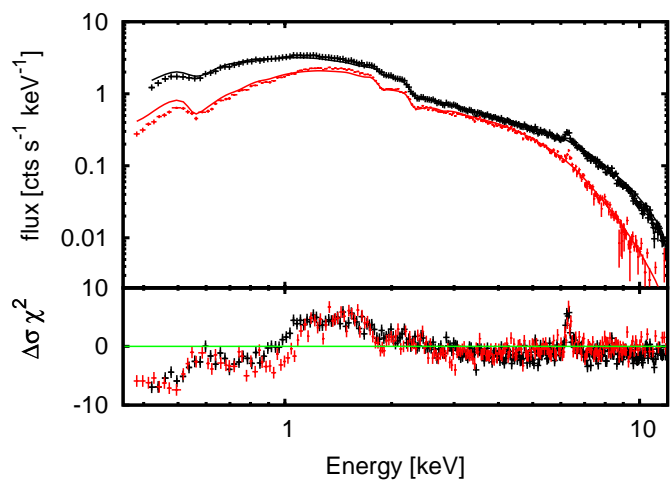} 
\end{tabular}
\caption{XMM-Newton PN (black) and joint MOS (red) spectrum of IRAS05078+1626 described 
by a simple power law model absorbed by Galactic neutral hydrogen in the line of sight 
with $n_{\rm H}=0.188 \times 10^{22}$\,cm$^{-2}$. The photon index of the power law is $\Gamma = 1.49$. 
The model reveals an apparent excess at $E=6.4$\,keV associated with the iron line K$\alpha$ 
and some wiggle-like residuals at lower energies. A more detailed view of the data residuals 
in these parts of the spectrum is shown in Fig.~\ref{pm_resid}.}
\label{powerlaw}
\end{figure}

\subsection{Mean spectral properties}
\label{msp}

The signal-to-noise ratio is very good up to high energies (Fig.~\ref{pndata_back}), 
so we fit the EPIC spectra spectra in the full energy range where they are
well calibrated (0.35--12\,keV).
The X-ray continuum is described by a power law model at energies above 2~keV, 
although the iron line at $E=6.4$\,keV is present (Fig.~\ref{powerlaw}). 
The photon index of the power law is $\Gamma \simeq 1.49(1)$.\footnote{All presented 
errors represent 90~$\%$ confidence level for a single interesting parameter, 
and the errors quoted in brackets are related to the last digit in the number.}
In this and all subsequent models we included absorption by Galactic gas
matter along the line of sight with column density $n_{\rm H}=0.188 \times 10^{22}$\,cm$^{-2}$. 
This value is from the Leiden/Argentine/Bonn H\,I measurements \citep{2005A&A...440..775K}. 
We used the {\tbabs} model \citep{wilms00} to fit the absorption produced by the Galactic interstellar matter. 

\begin{figure*}[bht]
\includegraphics[width=0.48\textwidth]{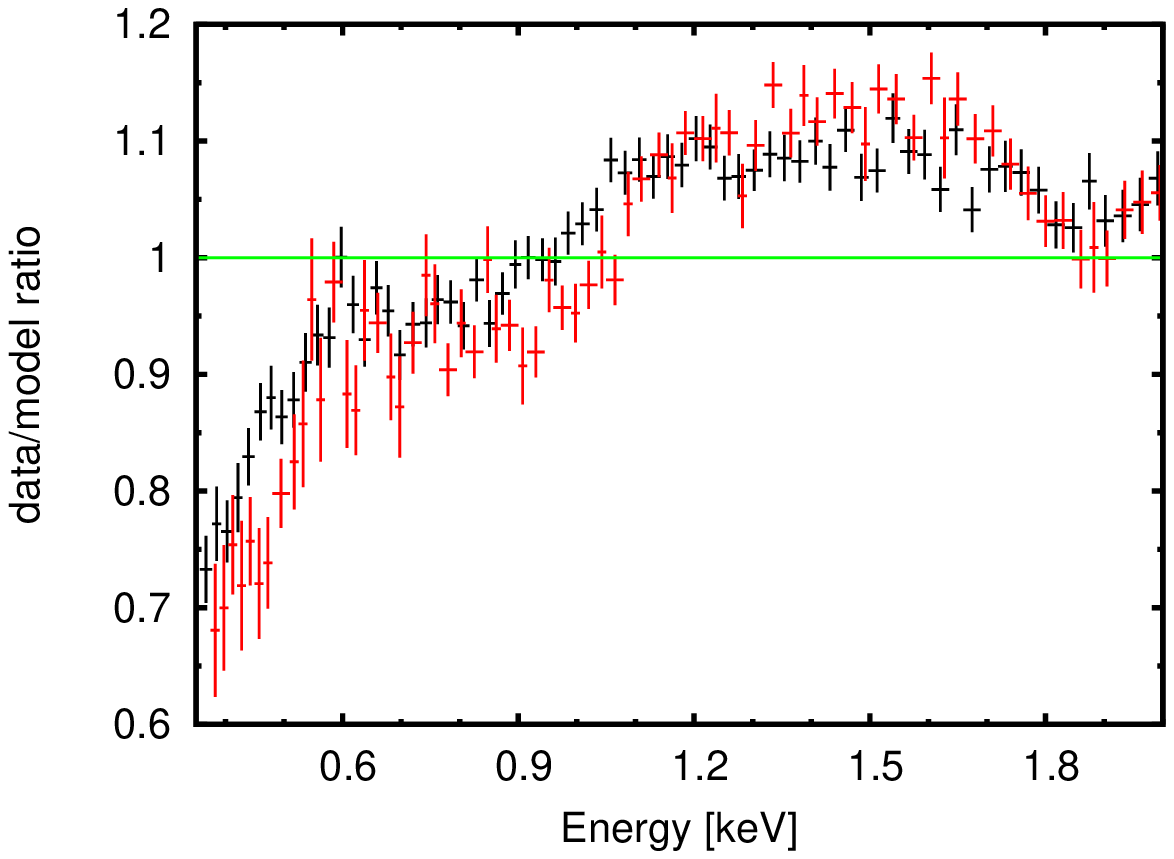}
\includegraphics[width=0.48\textwidth]{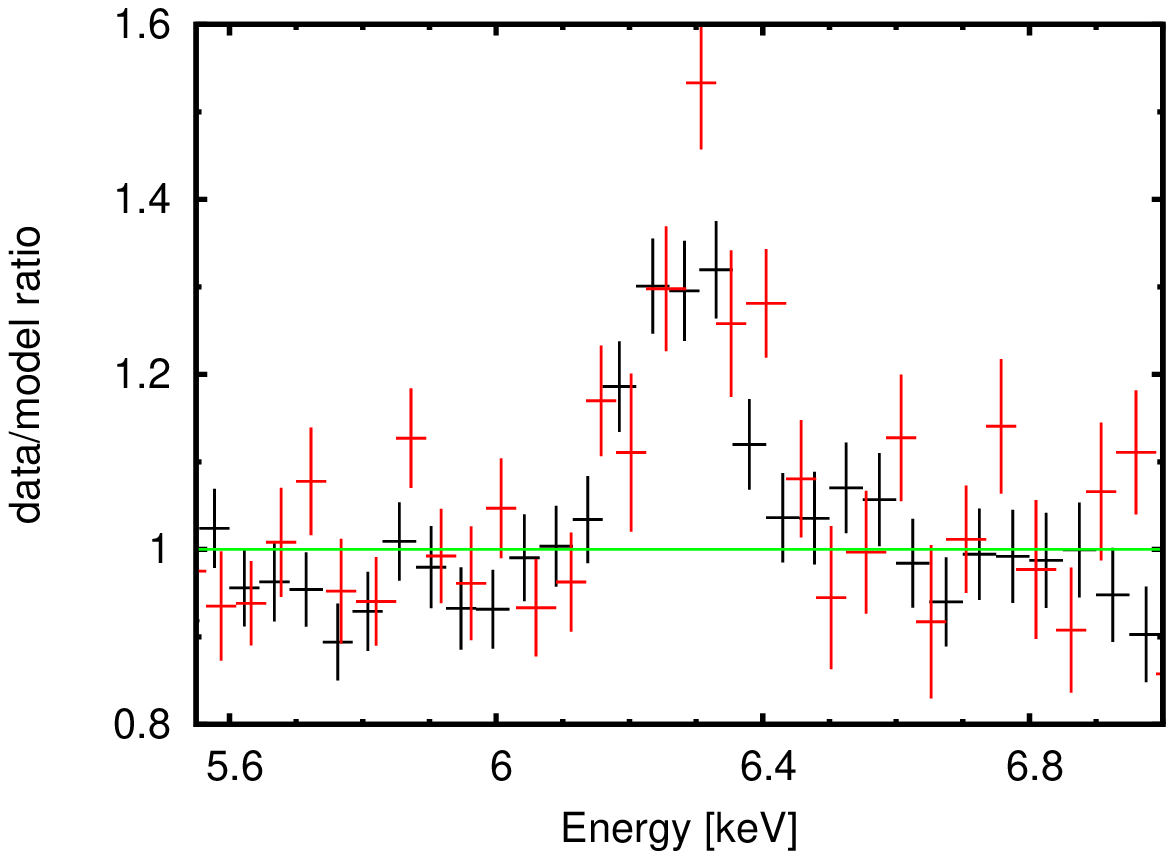}
\caption{Ratios of the simple power law model (the same as in Fig.~\ref{powerlaw}) to the data
in different energy bands: \textbf{left:} at lower energies, 
\textbf{right:} in the iron line band, where the narrow K$\alpha$ line
 at the rest energy $E=6.4$\,keV is prominent (observed at $E=6.29$\,keV due to
the cosmological redshift). Black crosses correspond to the PN data points,
while the red crosses correspond to the MOS data points.}
\label{pm_resid}
\end{figure*}

We applied the simple {\tbabs}*{\powerlaw} model to both PN and MOS spectra.
The $\chi^{2}$ value is 3557 with 528 degrees of freedom ($\chi^{2}/\nu=6.7$) 
in the $0.35-12.0$\,keV energy range. 
The spectra differ from the power law model not only around $E=6.4$\,keV
but also at lower energy band $0.35-2.0$\,keV (adding a Gaussian line model
to fit the iron line improves the fit only to $\chi^{2}/\nu = 3384/524 \doteq 6.5$). 
Residuals against this model are shown in Fig.~\ref{pm_resid}. 

The residuals at lower energies are usually attributed to a warm absorber, 
i.e.\ absorption by totally or partially ionised matter; see, e.g.,
\citet{2003ApJ...599..933N,blustin05,2007ApJ...659.1022K} for more 
information about warm absorbers in Seyfert galaxies.
The spectrum can be also affected by the so-called soft excess,
which can be either caused by reflection by the ionised surface of the accretion 
disc \citep{2006MNRAS.365.1067C} or by partially ionised 
and Doppler smeared absorption \citep{2006MNRAS.371L..16G}. 

The spectral residuals reveal certain discrepancies between the PN and MOS spectra
(see Fig.~\ref{pm_resid} with the data/model ratios of both spectra, PN and MOS, 
with the identical model parameters of the spectra). The level
of discrepancy is, however, comparable to the level of systematic uncertainties
in the cross-calibration between the EPIC cameras \citep{2006ESASP.604..937S}.
Nonetheless, we conservatively analyse the EPIC spectra separately. 
We use the same models for both spectra but allow the values 
of the model parameters to be different. 
The values of the photon index using the simple power law model differ from each other 
when fitting the spectra independently, resulting in a harder PN spectrum with $\Gamma = 1.60(1)$ 
compared to the MOS spectrum with $\Gamma = 1.54(1)$, ignoring the energies 
below 2\,keV and also between 5.5-7.5\,keV.
Although the absolute value of these spectral index
measurements does not have a direct physical meaning,
given the simplicity of the model applied on a small 
energy band, the comparison between them is illustrative 
of the quality of the cross-calibration between the EPIC cameras.
Differences of the order of $\Delta \Gamma \simeq$ 0.06 
in the hard X-ray band are consistent with current systematic 
uncertainties \citep{2006ESASP.604..937S}.

\subsection{RGS spectrum}
\label{rgs}

We jointly fit the unbinned first-order spectra of the two RGS cameras 
with the same model's parameter values except the overall normalisations.
The continuum is well-fitted by the simple power law model with the photon index $\Gamma = 1.57$.
We searched further for narrow emission and absorption lines in the spectrum using
several {\zgauss} models with the intrinsic width $\sigma$ set to zero. We calculated the 90\,\% confidence 
interval for a blind search, as $P=P_{0}/N_{\rm trial}$, where $N_{\rm trial}=N_{\rm bins}/3 = 3400/3$
and $P_{0}=0.1$. For the RGS data $P \doteq 8.8\times 10^{-5}$, to which $\Delta C = 22.4$
corresponds assuming the Student probability distribution.
The only line fulfilling this criterion by improving the fit about $\Delta C=31.7$ 
is an emission line at the energy $E=0.561 \pm 0.001$\,keV ($22.10 \pm 0.04$\,\AA)
and the equivalent width $7^{+5}_{-3}$\,eV.
We identify it with the forbidden line of the O~VII triplet ($E_{\rm LAB} = 0.561$\,keV).

\subsection{EPIC spectrum}
\label{res}

The forbidden line of the OVII triplet is clear signature of a photoionised plasma. 
No significant features were detected that may be expected alongside the O~VII~(f) line, 
if it were produced in a collisionally ionised plasma, such as the resonance 
line in the OVII triplet or the OVIII~Ly$\alpha$. 
This led us to try and explain the residuals against a power law model
in the soft X-rays as effect of intervening ionised absorption gas.
We used the {\xstar} model version 2.1ln7c \citep{2001ApJS..133..221K}\footnote{http://heasarc.gsfc.nasa.gov/docs/software/xstar/xstar.html} 
to calculate a grid of tabular models with the input parameters constrained 
from the preliminary data analysis with simple models whenever possible
(photon index $\Gamma \approx 1.7$, density $\rho \leq 10^{14}$\,cm$^{-3}$,
luminosity $L \leq 10^{44}$\,erg\,s$^{-1}$, column density 
$10^{19}$\,cm$^{-2} \leq n_{\rm H} \leq 10^{25}$\,cm$^{-2}$, and ionization parameter
$-5 \leq \log \xi \leq 5$).

\begin{figure}[tb!]
\begin{center}
\includegraphics[width=0.49\textwidth]{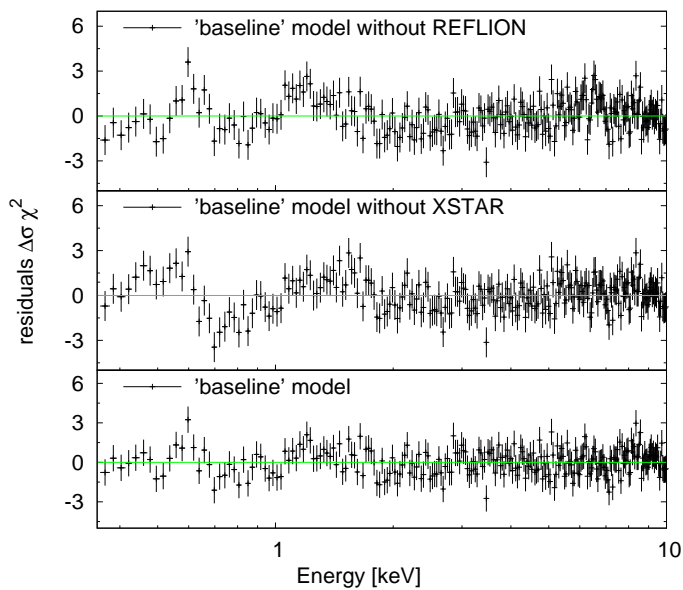}
\caption{Residuals of the PN data from the `baseline' model
including both ionised reflection and absorption (\textbf{lower}),
only reflection (\textbf{middle}), and only absorption (\textbf{upper}).}
\label{del_chi}
\end{center}
\end{figure}

\begin{table*}[tb!]
\caption{Parameters of the `baseline' and `final' models.}
\begin{center}
\begin{tabular}{c|c|c|c|c|c|c} 
	\hline \hline \rule{0cm}{0.5cm}
 Model 	&	Model		&\multicolumn{2}{c|}{`baseline' model}	&	\multicolumn{3}{c}{`final' (`double reflection') model}	 \\
  component	&	 parameter	&	PN	&	MOS &	PN	&	MOS   &  PN+MOS+RGS  \\
	\hline
\rule[-0.7em]{0pt}{2em} \zphabs	& $n_{\rm H} [10^{22}$\,cm$^{-2}]$	& $0.104^{+0.005}_{-0.007}$	&	$0.129^{+0.007}_{-0.007}$	&	$0.102^{+0.009}_{-0.005}$	& $0.120^{+0.008}_{-0.005}$	& $0.106^{+0.004}_{-0.004}$ \\
\hline
\rule[-0.7em]{0pt}{2em} \xstar	& $n_{\rm H} [10^{22}$\,cm$^{-2}]$	&	$130^{+20}_{-10}$	& 	$170^{+20}_{-20}$	&	$120^{+30}_{-30}$	&	$150^{+70}_{-20}$ 	&	$130^{+20}_{-20}$	\\
\rule[-0.7em]{0pt}{2em}	&	$\log\,\xi$			&$2.3^{+0.1}_{-0.1}$	&	$2.4^{+0.1}_{-0.1}$	&	$2.2^{+1.4}_{-0.6}$	&	$2.5^{+1.0}_{-0.5}$  &	$2.5^{+1.0}_{-0.4}$	\\
\rule[-0.7em]{0pt}{2em}	&	He/He$_{\rm Solar}$ - Ca/Ca$_{\rm Solar}$ 	&	$1$ (f)	
&	$1$ (f)	&	$1$ (f)	&	$1$ (f) &	$1$ (f)	\\
\rule[-0.7em]{0pt}{2em}	&	Fe$/$Fe$_{\rm Solar}$ - Ni$/$Ni$_{\rm Solar}$		&	$0.2^{+0.1}_{-0.2}$	&	$0.1^{+0.1}_{-0.1}$		&	$1.2^{+0.3}_{-0.3}$	&	$0.9^{+0.2}_{-0.2}$	&	$1.1^{+0.2}_{-0.2}$	\\
\hline
\rule[-0.7em]{0pt}{2em} \powerlaw &	$\Gamma$		&	$1.81^{+0.03}_{-0.05}$		&	$1.80^{+0.05}_{-0.05}$	&	$1.75^{+0.10}_{-0.03}$		&	$1.74^{+0.07}_{-0.03}$	&	$1.76^{+0.04}_{-0.02}$	\\
\rule[-0.7em]{0pt}{2em}	&	normalisation		&	$\left(7 \pm 1\right) \times 10^{-4}$		&	$(7\pm1) \times 10^{-4}$	&	$(6\pm1) \times 10^{-4}$		&	$(7\pm2) \times 10^{-4}$	& .....	\\
\hline
\rule[-0.7em]{0pt}{2em} \reflion &	$\Gamma$	&	$1.81$ (b)	&	$1.80$ (b)	&	$1.75$	(b)	&	$1.74$ (b)	&	$1.76$ (b)	\\
\rule[-0.7em]{0pt}{2em}	&	$\log\,\xi$		&	$3.0^{+0.2}_{-0.2}$		&	$3.2^{+0.2}_{-0.2}$	&	$3.0^{+0.2}_{-0.2}$		&	$3^{+2}_{-3}$	&	$3.0^{+0.1}_{-0.2}$	\\
\rule[-0.7em]{0pt}{2em}	&	Fe$/$Fe$_{\rm Solar}$ - Ni$/$Ni$_{\rm Solar}$	& 	$0.2$ (b)	&	$0.1$ (b)	&	$1.2$ (b)	&	$0.9$ (b)	&	$1.1$ (b)	\\
\rule[-0.7em]{0pt}{2em}	&	normalisation 		&	$(3 \pm 2) \times 10^{-9}$	&	$(3 \pm 2) \times 10^{-9}$	&	$(2\pm1) \times 10^{-9}$	&	$(1\pm1) \times 10^{-9}$		& .....		\\
\hline
\rule[-0.7em]{0pt}{2em} \reflion~2&	normalisation 		&	-	&	-	&	$3\pm1 \times 10^{-7}$	&	$4\pm1 \times 10^{-7}$		& .....		\\
\hline
\rule[-0.7em]{0pt}{2em} \zgauss &	E [keV]		&	$6.40^{+0.01}_{-0.01}$	&	$6.44^{+0.02}_{-0.02}$	& -  & -  & -	\\
\rule[-0.7em]{0pt}{2em}	&	$\sigma$ [keV]		&	$0.06^{+0.03}_{-0.04} $	&	$0.02^{+0.04}_{-0.02} $	& -  & - & - \\
\rule[-0.7em]{0pt}{2em}	&	z	&	$0.018$	(f)	&	$0.018$	(f)  & - & - & -\\
\rule[-0.7em]{0pt}{2em}	&	normalisation 		&	$(3\pm1) \times 10^{-5}$ &	$(3\pm1) \times 10^{-5}$	& -  & - & -	\\
\hline
\hline
\rule[-0.7em]{0pt}{2em} $\chi^2/\nu$	& &	246/264		&	405/243		&	256/266		&	404/244		
&	$C/\nu$ = 1551/1347  \\
\end{tabular}
\end{center}
{
Note: The sign (f) after a value means that the value was fixed during the fitting procedure. 
The sign (b) means that the parameter value was bound to the value of the corresponding 
parameter of the previous model component. The sign ``-'' means that the model component
is not included in the total model, while dots in the right column only mean that there
are more values related to the individual spectra which are not necessary to be all 
shown in the table.
}
\label{model}
\end{table*} 

A single-zone warm absorber component modifying the power law continuum dramatically
improved the fit from $\chi^2 / \nu = 1850/270 \doteq  6.9$ 
to $\chi^2 / \nu = 402/270 \doteq  1.5$ for the PN spectrum.
The ionization parameter converged to a very low value, and we found 
that this almost neutral absorption can be successfully reproduced with {\zphabs}, 
which is a simpler model than {\xstar}, so we preferred this possibility.
The addition of another warm absorber zone improves the fit 
to $\chi^2 / \nu = 320/266 \doteq 1.2$ for the PN spectrum, and it requires
the ionization parameter $\log \xi \cong 3.9$.
We checked that adding another warm absorber zones does not improve
the fit significantly.
The residuals from the model (see Fig.~\ref{del_chi}, upper panel)
reveal an extra emission that remains at low energies, as well as 
around the iron K~$\alpha$ line band.

These features can come from reflection of the primary radiation on the surface 
of the accretion disc, so we added the {\reflion} model \citep{2005MNRAS.358..211R}, 
which calculates the ionised reflection for an optically thick atmosphere with constant density.
We examined the significance of the addition of the reflection component 
into the complex model of the PN spectrum by the statistical F-test. 
The low value of the F-test probability ($5 \times 10^{-15}$) strongly favours this additional model component.
The best fit was now $\chi^2/\nu = 246/264 \doteq 0.95$ for the PN spectrum.
We hereafter call this model the `baseline' model; in the XSPEC notation:
{\tbabs} $\times$ {\zphabs}$_{\,\rm N}$ $\times$ {\xstar} $\times$ ({\powerlaw} + {\reflion} + {\zgauss}).

The parameter values of the `baseline' model are presented in the Table~\ref{model}.
The quoted errors of the parameters represent a 90~$\%$ confidence 
level for a single interesting parameter. The measurement is obviously 
affected by a much larger systematic error, which, however, 
could be properly quantified only if we knew the ``right'' model.
The value of the power law photon index increased to $\Gamma \approx 1.8$
compared to the simple model applied to the data in Sect.~\ref{msp},
because we included of the additional local absorption in the model.
The data residuals from the model are shown in the lower panel of Fig.~\ref{del_chi}.
In the same figure, we also show residuals from best fit
performed with the `baseline' model, excluding the ionised absorption
(middle panel) and the ionised reflection component (upper panel).

The narrow iron K~$\alpha$ line with the rest energy $E=6.40 \pm 0.01$\,keV,
the width $\sigma=0.06 \pm 0.03$\,keV, and the equivalent width $EW = 82 \pm 15$\,eV
evidently represents cold reflection. This suggests an origin of this spectral component 
in the outer part of the disc, or from the torus. 
The cold reflection is also supposed to contribute to the soft part of the spectrum
with the individual emission lines. For this reason, we replaced the Gaussian profile
in the `baseline' model with another {\reflion} component (called as {\reflion}~2 
in the Table ~\ref{model}) with the same values for the photon index 
and abundances as the {\reflion}~1 model component.
The ionization parameter was kept free during the fitting procedure, but 
it very quickly converged to its lowest value $\xi = 30$ ($\log \xi = 1.477$). 
The advantage of the {\reflion} model compared to the other available 
reflection models is that it also includes the soft X-ray lines, 
with the disadvantage in this case that the ionization parameter cannot be set to zero. 

\begin{figure}[tb!]
\begin{center}
\includegraphics[width=0.48\textwidth]{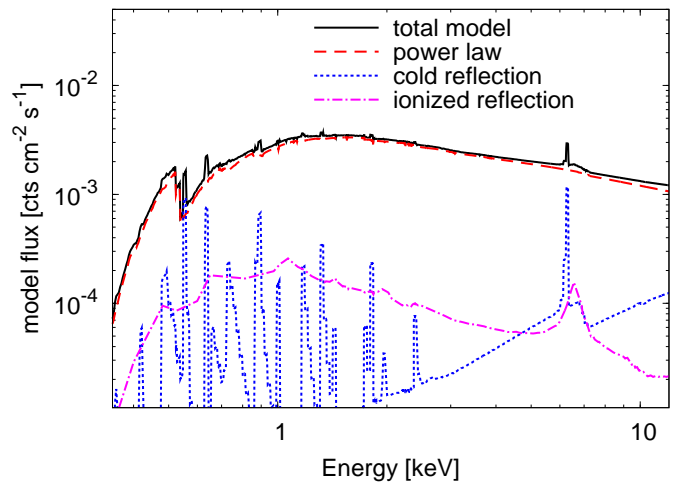}
\caption{The `final' model. The total model is shown in black (solid line), the primary radiation
is red (dashed), the {\reflion} components are blue (dotted) for cold reflection 
and magenta (dot-dashed) for ionised reflection. 
}
\label{emo}
\end{center}
\end{figure}

This `double reflection' model, in the XSPEC notation
{\tbabs} $\times$ {\zphabs}$_{\,\rm N}$ $\times$ {\xstar} $\times$ ({\powerlaw} + {\reflion} + {\reflion}),
does not significantly improve the fit goodness over the `baseline' model 
(with $\chi^2/\nu = 256/265 \doteq 0.96$ for the PN spectrum), 
but it represents a more self-consistent astrophysical picture.
Therefore, we call the `double reflection' model as `final' model.
In contrast to the `baseline' model, it does not require subsolar iron abundances, 
see the Table~\ref{model}, where the parameter values for this model are presented.
The `final' model with each component separately drawn is shown in Fig.~\ref{emo}.
All the plotted components are absorbed by a warm absorber surrounding 
the central accretion disc and two kinds of cold absorber -- one from 
Galactic interstellar matter and one from local absorber in the host galaxy.

\begin{figure}[tb!]
\begin{center}
\includegraphics[width=0.48\textwidth]{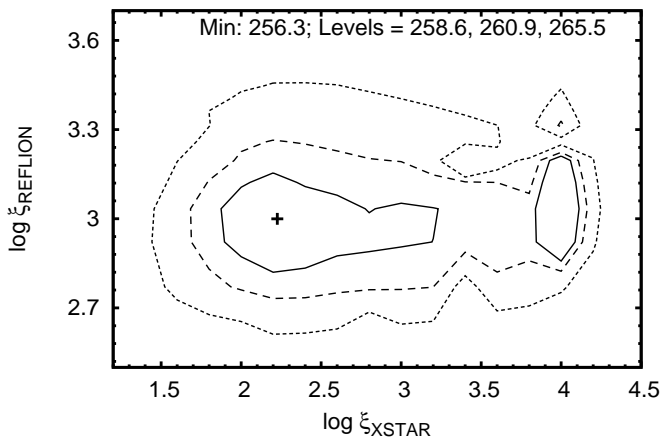}
\caption{The contour plot of the ionization parameters of the {\reflion} model,
representing the ionised accretion disc, and of the {\xstar} model, representing
the warm absorber in the `final' model. 
The individual curves correspond to the 1\,$\sigma$, 2\,$\sigma$, and 3\,$\sigma$
levels. The position of the minimal value of $\chi^{2}$ found by the fitting 
procedure is marked by a cross. The corresponding $\chi^2$ values are given
in the plot.}
\label{cont_ion}
\end{center}
\end{figure}

\begin{table}[tb!]
\caption{Flux values of the `final' model and its individual components.} 
\begin{center}
\begin{tabular}{c|cc|cc} 
	\hline \hline
\rule[-0.6em]{0pt}{1.7em} Model	&	\multicolumn{2}{c|}{Flux at $0.5-2\,$keV} &	\multicolumn{2}{c}{Flux at $2-10\,$keV}	\\
\rule[-0.6em]{0pt}{1.7em} component	&	\multicolumn{2}{c|}{$[10^{-12}$erg\,cm$^{-2}$\,s$^{-1}$]}	&	\multicolumn{2}{c}{$[10^{-12}$erg\,cm$^{-2}$\,s$^{-1}$]}	 \\
\rule[-0.6em]{0pt}{1.7em} 	& 	PN	&	MOS &	PN	&	MOS\\
\hline
\rule[-0.6em]{0pt}{1.7em} total	model	& $7.05^{+0.03}_{-0.03}$	&	$7.00^{+0.03}_{-0.03}$	& $25.0^{+0.1}_{-0.2}$	&	$25.4^{+0.1}_{-0.2}$	\\
\rule[-0.6em]{0pt}{1.7em} unabsorbed model	& $16.6^{+0.2}_{-0.2}$	&	$16.7^{+0.2}_{-0.2}$	& $25.7^{+0.2}_{-0.2}$	&	$26.5^{+0.2}_{-0.2}$	\\
\rule[-0.6em]{0pt}{1.7em} {\powerlaw}	& $13.6^{+0.1}_{-0.2}$	&	$14.3^{+0.1}_{-0.1}$	& $22.6^{+0.3}_{-0.2}$	&	$23.8^{+0.1}_{-0.1}$	\\
\rule[-0.6em]{0pt}{1.7em} {\reflion}$_{\rm ion}$	& $1.9^{+0.2}_{-0.2}$	&	$1.1^{+0.1}_{-0.1}$	& $1.6^{+0.1}_{-0.2}$	&	$0.9^{+0.1}_{-0.1}$	\\
\rule[-0.6em]{0pt}{1.7em} {\reflion}$_{\rm cold}$	& $1.1^{+0.1}_{-0.1}$	&	$1.3^{+0.1}_{-0.1}$	& $1.5^{+0.1}_{-0.1}$	&	$1.8^{+0.2}_{-0.1}$	\\
\hline
\rule[-0.6em]{0pt}{1.7em} $R_{\rm ion}$ $^*$		& $0.12^{+0.01}_{-0.01}$	&	$0.07^{+0.01}_{-0.01}$ & $0.06^{+0.01}_{-0.01}$	&	$0.03^{+0.01}_{-0.01}$\\
\rule[-0.6em]{0pt}{1.7em} $R_{\rm cold}$ $^*$	& $0.07^{+0.01}_{-0.01}$	&	$0.08^{+0.01}_{-0.01}$ & $0.06^{+0.01}_{-0.01}$	&	$0.07^{+0.01}_{-0.01}$\\
\end{tabular}
\end{center}

{$^*$ the ratios of the reflection component flux values to the flux 
value of the total unabsorbed model (sum of the primary and reflected radiation).}
\label{rratio}

\end{table} 

Some model parameters were not allowed to vary during the fitting procedure. 
The redshift of the ionised absorber was fixed to the source cosmological value, 
because leaving it free yields a negligible improvement in the quality of the fit.
Second, we used the same iron abundances across all the components in the model.

In the `final' model, the warm absorber ionization parameter is consistent
with the ionised reflection component.
This result is also presented in Fig.~\ref{cont_ion}, 
where the contour lines related to the $1\sigma$, $2\sigma$, and $3\sigma$ levels 
of $\chi^{2}$ between the ionization parameters of the two model components are present.

Table~\ref{rratio} summarises flux values of the individual components of the `final' model
for both PN and MOS spectra for two energy bands, $0.5-2$\,keV and $2-10$\,keV,
and also shows fractions of the reflection radiation to the total emission (sum of the
primary and reprocessed radiation). The flux ratio is almost equally 
shared between the cold and ionised reflection components, 
and its value is in total $R<0.2$.
The absorption--corrected luminosity values of the source in the same energy bands are
$L\,_{0.5-2\rm\,keV}=(1.21 \pm 0.02)\times10^{43}$\,erg\,s$^{-1}$
and $L\,_{2-10\rm\,keV}=(1.87 \pm 0.02)\times10^{43}$\,erg\,s$^{-1}$, respectively.

\begin{figure}[tb!]
\begin{center}
\includegraphics[width=0.48\textwidth]{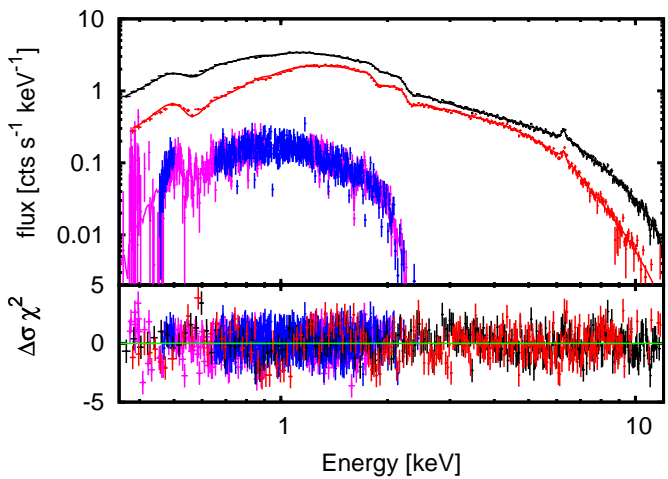}
\caption{The joint fit of all spectra of the XMM-Newton instruments -- 
PN (black), MOS (red), RGS\,1 (magenta), and RGS\,2 (blue), together with
the model residuals.}
\label{joint}
\end{center}
\end{figure}

We also used the `final' model for a joint fitting of all 
the XMM-Newton instruments (PN, MOS and both RGS spectra) together.
The parameter values were bound among all the spectra, only normalisation factors were allowed to vary.
The goodness of the joint fit is given in C-statistics because the RGS data are unbinned 
(and each individual bin contains only a few counts). The result is
$C = 1551$ for a number of degrees of freedom $\nu = 1347$.
All the spectra, together with the residuals, are shown in Fig.~\ref{joint},
and the corresponding parameters in the last column of Table~\ref{model}.

\section{Discussion}
\label{discussion}

\subsection{Constraints on the location of the absorbers}
\label{geometry}

In this section we discuss a possible location of the absorber's system
in the `final' model. Photoelectric absorption is almost invariably
observed in Type~2 Seyferts \citep{1991ApJ...366...88A,1997ApJS..113...23T,2002ApJ...571..234R}
and generally attributed to an optically thick matter responsible for orientation-dependent 
classification in AGN unification scenarios \citep{1985ApJ...297..621A,1993ARA&A..31..473A}.

Because the IRAS 05078+1626 galaxy is probably viewed under an intermediate 
inclination between unobscured Seyfert 1s and obscured Seyfert 2s, 
the torus rim may also intercept the line of sight to the AGN
and absorb the radiation coming from the centre.
The cold absorption can, however, also be associated with the interstellar matter 
of the galaxy \citep{2006A&A...449..551L}.

Both reflection components are inside the ionised absorber in the `final' model.
The geometrical interpretation is that the cold reflection occurs on the outer
parts of the disc or the inner wall of the torus. Reflection on the nearer 
part of the torus is heavily absorbed by the torus itself, so only radiation
reflected on the farther peripheral part of the torus can reach the observer
after passing through the warm absorber.
However, an alternative scenario, in which the cold reflection is unaffected by the warm absorber, i.e.,
{\tbabs} $\times$ {\zphabs} $\times$ [{\reflion}$_{\rm cold}$ + {\xstar} $\times$ ({\powerlaw} + {\reflion}$_{\,\rm disc}$ )],
is also acceptable with  $\chi^{2}/\nu = 265/265$.

The lack of constraints on the variability in the warm absorbed features 
\citep{2007ApJ...659.1022K}, caused by the moderate dynamical range of the primary continuum, 
as well as statistical limitations in our spectra, prevents us from precisely 
constraining the location of the warm absorber.

\begin{figure}[tb!]
\begin{center}
\includegraphics[width=0.48\textwidth]{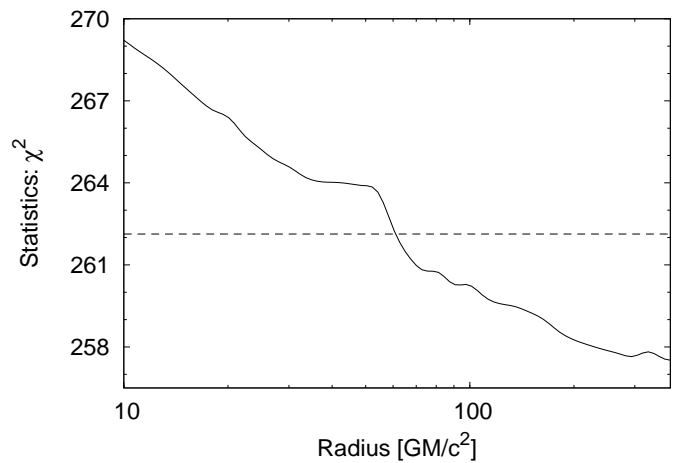}
\caption{The best-fit values of $\chi^2$ statistics for the inner disc radius parameter,
which we obtained by gradually stepping it from the horizon radius 
to the outer radius of the disc ($400 R_{\rm g}$). 
The dashed line is the 90\% confidence level for one interesting parameter. 
}
\label{kyconvlevel}
\end{center}
\end{figure}

\subsection{Constraints on the location of the ionised reflector}
\label{disc}

The ionised reflection might occur either at the inner wall of a warm absorber cone
or on the accretion disc. Even in the latter case, the reflection cannot occur
arbitrarily close to the black hole. In this section we investigate 
the constraints of the accretion disc location and structure, which can be drawn 
from the lack of the significant relativistic blurring of the disc reflection component.

We convolved the ionised reflection component with the 
fully relativistic {\kyconv} model \citep{2004ApJS..153..205D}.
Two assumptions about the disc emissivity were considered.
First, the radial part of the intensity decreases with the power 
of the disc radius $q$ ($I \propto r^{-q}$),
where the value of $q$ was allowed to vary between $2$ and $3.5$. 
Second, the angular dependence was assumed to be isotropic
which seems to be appropriate approximation for our situation
of an X-ray irradiated accretion disc \citep{2009A&A...507....1S}.

We examined the expected confidence levels of the best-fit values
of the disc's inner radius
by stepping this parameter in the whole range of its possible
values -- from the horizon to the outer disc radius, which we set to 400
gravitational radii ($R_{\rm g} \equiv GM/c^{2}$). 
The results are shown in Fig.~\ref{kyconvlevel}. At the 90~\% confidence level,
the accretion disc is not allowed to extend closer to the black hole than 
$60 R_{\rm g}$. 

The ``relativistic blurring method'' would be less appropriate
in looking for the imprints of the innermost parts of the accretion disc
if the disc were too highly ionised (log$~\xi \approx 4$) and the narrow
reflection features were not present \citep{2005MNRAS.358..211R}. However,
the ionization parameter value of the reflection component is not so high
in the `final' model and the dominant feature is the intermediately 
ionised iron line ($E \approx 6.7$\,keV). If we assumed a stratified 
disc with the ionization state decreasing with the radius from the centre, 
the hydrogen-like iron line would be also expected to appear in the spectrum
(as an intermediate stage between the over-ionised and mildly ionised contribution).
Because it is not detected in the data,
the accretion disc truncation provides a more reasonable explanation
of missing signatures of the relativistic blurring.

\subsection{Mass accretion rate}

Disc truncation is expected in low--luminosity AGN where the inner
accretion flow is advection-dominated \citep[and references therein]{1994ApJ...428L..13N, 
1997ApJ...489..865E,2008NewAR..51..733N}.
The transition from the outer standard accretion disc may occur, e.g., via
the disc evaporation mechanism \citep{2000A&A...361..175M, 2009ApJ...707..233L}.
The observational evidence of a truncated accretion disc
in low--luminosity AGN was reported e.g. by \citet{1996ApJ...462..142L, 1999ApJ...525L..89Q}. 
However, its presence is also suggested in some observations of Seyfert galaxies 
\citep{2000ApJ...537L.103L, 2000ApJ...536..213D, 2003ApJ...586...97C, 2009ApJ...705..496M}
and even a quasar \citep{2005A&A...435..857M} where the luminosity value is estimated as a half
of the Eddington value. Generally, it is expected that the lower the luminosity, $L/L_{\rm Edd}$,
the larger transition radius \citep[see][and references therein]{2004ApJ...612..724Y}.
Furthermore, we investigate whether the disc truncation hypothesis is consistent 
with the IRAS~05078+1626 luminosity. To have these quantities in Eddington units, 
we first estimated the mass of the black hole. 

IRAS~05078+1626 belongs to the sample of the
infrared-selected Seyfert 1.5 galaxies observed by a 2.16~m optical telescope 
\citep{2006ApJ...638..106W} where, among others, the velocity dispersion 
in the O~III emission line was measured. 
The correlation between the O~III line width and the mass of active galactic
nucleus was discussed in \citet{2000ApJ...544L..91N} and \citet{2003ApJ...585..647B}. 
The value from the optical measurements, $\sigma_{\rm O\,III} \approx 130$\,km\,s$^{-1}$, 
corresponds to the mass $M \approx 4\times10^{7} M_{\odot}$ using 
a correlation plot in \citet{2003ApJ...585..647B}. The scatter of the correlation
is somewhat large with the reported limit of a factor of 5 for an uncertainty
in the black hole mass determination, so the value only provides 
an order of magnitude estimation.

The value of the Eddington luminosity is
$L_{\rm Edd} \doteq 1.3 \times 10^{38} M/M_{\odot} $\,erg\,s$^{-1} \approx 5 \times 10^{45}$\,erg\,s$^{-1}$
for the given value of the mass.
We used luminosity-dependent corrections by \citet{2004MNRAS.351..169M} to estimate
the bolometric luminosity of IRAS~05078+1626 from the X-ray luminosity. Its value is 
$L \approx 5 \times 10^{44}$\,erg\,s$^{-1} \approx 10^{-1} L_{\rm Edd}$.
Correspondingly, the mass accretion rate, $\dot{M} = L/c^2$,
is sub-Eddington with $\dot{M} \approx 0.1 \dot{M}_{\rm Edd}$.
This value is typical of less luminous Seyfert galaxies \citep[see for example][]{2009A&A...495..421B},
and is consistent with the disc truncation hypothesis.


\section{Conclusions}
\label{conclusions}

The X-ray continuum spectrum of the Seyfert galaxy IRAS 05078+1626 
is dominated by a power law with a standard value of the photon index 
($\Gamma \cong 1.75$ in the `final model'). 
The residuals from the power law continuum can be interpreted in terms 
of the warm absorber surrounding the accretion disc,
and the reflection of the primary radiation from the ionised matter and on the
cold torus. The outgoing radiation is absorbed by cold matter 
($n_{\rm H} \approx 1 \times 10^{21}$\,cm$^{-2}$), which can be either 
located in the inner side of the torus or caused by gas in the host galaxy. 
The type of the galaxy determined from the previous infrared and optical 
research is Seyfert 1.5, suggesting that the active nucleus 
could be seen at large inclination, consistent with either interpretation 
or even allowing a combination of both.

The ionised warm absorber occurs in the central part of the AGN.
Its column density was found to be $n_{\rm H} \geq 1 \times 10^{24}$\,cm$^{-2}$,
which is a rather high value compared to the warm absorbers detected
in the other Seyfert galaxies \citep{blustin05}.
This may be because we are looking through a longer
optical path of a conical non-relativistic outflow due to the high
inclination of the system.
The ionization parameter of the warm absorber is $\log \xi_{\rm WA} = 2.5 \pm 1.0$,
which is comparable to the value related to the ionised reflection
$\log \xi_{\rm reflection} = 3.0 \pm 0.2$, suggesting a link between them.

If the ionised reflection is associated to the warm absorber
(e.g. the inner walls of a conical outflow), the lack of spectral
features associated with the accretion disc is a natural consequence thereof.
If, instead, the ionised reflection occurs at the accretion disc, 
it cannot extend up to the marginally stable orbit. 
The lack of the significant relativistic blurring of this model component
requires the disc to be truncated (inner disc radius $R_{\rm in} \geq 60\,R_{g}$).
This idea is also supported by the low ratio of the reflection radiation 
to the primary one, $R < 0.2$, and also by the relatively low mass-accretion 
rate $\dot{M} \approx 0.1\dot{M}_{\rm Edd}$ determined
from the source luminosity.

\begin{acknowledgements} 
The authors are grateful for useful comments and suggestions 
by Michal Dov\v{c}iak, Ren\'{e}~W.~Goosmann, and the
participants of the 3rd `FERO' (Finding Extreme Relativistic Objects) 
workshop, held in September 2009 in Rome.
JS acknowledges the support from the doctoral 
student programme of the Czech Science Foundation, ref.\ 205/09/H033,
and the research grant of the Charles University in Prague, ref.\ 33308.
VK appreciates the continued support from research grants of the 
Czech Science Foundation, ref.\ 205/07/0052,
and ESA Plan for European Cooperating States, project No.\ 98040.
\end{acknowledgements}

\bibliographystyle{aa} 
\bibliography{03} 

\begin{thebibliography}{52}
\expandafter\ifx\csname natexlab\endcsname\relax\def\natexlab#1{#1}\fi

\bibitem[{{Ajello} {et~al.}(2008){Ajello}, {Rau}, {Greiner}, {Kanbach},
  {Salvato}, {Strong}, {Barthelmy}, {Gehrels}, {Markwardt}, \&
  {Tueller}}]{ajello08}
{Ajello}, M., {Rau}, A., {Greiner}, J., {et~al.} 2008, \apj, 673, 96

\bibitem[{{Antonucci}(1993)}]{1993ARA&A..31..473A}
{Antonucci}, R. 1993, \araa, 31, 473

\bibitem[{{Antonucci} \& {Miller}(1985)}]{1985ApJ...297..621A}
{Antonucci}, R.~R.~J. \& {Miller}, J.~S. 1985, \apj, 297, 621

\bibitem[{{Arnaud}(1996)}]{1996ASPC..101...17A}
{Arnaud}, K.~A. 1996, in Astronomical Society of the Pacific Conference Series,
  Vol. 101, Astronomical Data Analysis Software and Systems V, ed. G.~H.
  {Jacoby} \& J.~{Barnes}, 17

\bibitem[{{Awaki} {et~al.}(1991){Awaki}, {Koyama}, {Kunieda}, {Takano},
  {Tawara}, \& {Ohashi}}]{1991ApJ...366...88A}
{Awaki}, H., {Koyama}, K., {Kunieda}, H., {et~al.} 1991, \apj, 366, 88

\bibitem[{{Bianchi} {et~al.}(2009{\natexlab{a}}){Bianchi}, {Guainazzi}, {Matt},
  {Fonseca Bonilla}, \& {Ponti}}]{2009A&A...495..421B}
{Bianchi}, S., {Guainazzi}, M., {Matt}, G., {Fonseca Bonilla}, N., \& {Ponti},
  G. 2009{\natexlab{a}}, \aap, 495, 421

\bibitem[{{Bianchi} {et~al.}(2009{\natexlab{b}}){Bianchi}, {Piconcelli},
  {Chiaberge}, {Bail{\'o}n}, {Matt}, \& {Fiore}}]{2009ApJ...695..781B}
{Bianchi}, S., {Piconcelli}, E., {Chiaberge}, M., {et~al.} 2009{\natexlab{b}},
  \apj, 695, 781

\bibitem[{{Blustin} {et~al.}(2005){Blustin}, {Page}, {Fuerst},
  {Branduardi-Raymont}, \& {Ashton}}]{blustin05}
{Blustin}, A.~J., {Page}, M.~J., {Fuerst}, S.~V., {Branduardi-Raymont}, G., \&
  {Ashton}, C.~E. 2005, \aap, 431, 111

\bibitem[{{Boroson}(2003)}]{2003ApJ...585..647B}
{Boroson}, T.~A. 2003, \apj, 585, 647

\bibitem[{{Brightman} \& {Nandra}(2008)}]{2008MNRAS.390.1241B}
{Brightman}, M. \& {Nandra}, K. 2008, \mnras, 390, 1241

\bibitem[{{Cappi} {et~al.}(2006){Cappi}, {Panessa}, {Bassani}, {Dadina},
  {Dicocco}, {Comastri}, {della Ceca}, {Filippenko}, {Gianotti}, {Ho},
  {Malaguti}, {Mulchaey}, {Palumbo}, {Piconcelli}, {Sargent}, {Stephen},
  {Trifoglio}, \& {Weaver}}]{2006A&A...446..459C}
{Cappi}, M., {Panessa}, F., {Bassani}, L., {et~al.} 2006, \aap, 446, 459

\bibitem[{{Cash}(1976)}]{1976A&A....52..307C}
{Cash}, W. 1976, \aap, 52, 307

\bibitem[{{Chiang} \& {Blaes}(2003)}]{2003ApJ...586...97C}
{Chiang}, J. \& {Blaes}, O. 2003, \apj, 586, 97

\bibitem[{{Crummy} {et~al.}(2006){Crummy}, {Fabian}, {Gallo}, \&
  {Ross}}]{2006MNRAS.365.1067C}
{Crummy}, J., {Fabian}, A.~C., {Gallo}, L., \& {Ross}, R.~R. 2006, \mnras, 365,
  1067

\bibitem[{{den Herder} {et~al.}(2001){den Herder}, {Brinkman}, {Kahn},
  {Branduardi-Raymont}, {Thomsen}, {Aarts}, {Audard}, {Bixler}, {den Boggende},
  {Cottam}, {Decker}, {Dubbeldam}, {Erd}, {Goulooze}, {G{\"u}del}, {Guttridge},
  {Hailey}, {Janabi}, {Kaastra}, {de Korte}, {van Leeuwen}, {Mauche},
  {McCalden}, {Mewe}, {Naber}, {Paerels}, {Peterson}, {Rasmussen}, {Rees},
  {Sakelliou}, {Sako}, {Spodek}, {Stern}, {Tamura}, {Tandy}, {de Vries},
  {Welch}, \& {Zehnder}}]{rgs}
{den Herder}, J.~W., {Brinkman}, A.~C., {Kahn}, S.~M., {et~al.} 2001, \aap,
  365, L7

\bibitem[{{Done} {et~al.}(2000){Done}, {Madejski}, \&
  {{\.Z}ycki}}]{2000ApJ...536..213D}
{Done}, C., {Madejski}, G.~M., \& {{\.Z}ycki}, P.~T. 2000, \apj, 536, 213

\bibitem[{{Dov{\v c}iak} {et~al.}(2004){Dov{\v c}iak}, {Karas}, \&
  {Yaqoob}}]{2004ApJS..153..205D}
{Dov{\v c}iak}, M., {Karas}, V., \& {Yaqoob}, T. 2004, \apjs, 153, 205

\bibitem[{{Esin} {et~al.}(1997){Esin}, {McClintock}, \&
  {Narayan}}]{1997ApJ...489..865E}
{Esin}, A.~A., {McClintock}, J.~E., \& {Narayan}, R. 1997, \apj, 489, 865

\bibitem[{{Gabriel} {et~al.}(2004){Gabriel}, {Denby}, {Fyfe}, {Hoar}, {Ibarra},
  {Ojero}, {Osborne}, {Saxton}, {Lammers}, \& {Vacanti}}]{2004ASPC..314..759G}
{Gabriel}, C., {Denby}, M., {Fyfe}, D.~J., {et~al.} 2004, in Astronomical
  Society of the Pacific Conference Series, Vol. 314, Astronomical Data
  Analysis Software and Systems (ADASS) XIII, ed. {F.~Ochsenbein, M.~G.~Allen,
  \& D.~Egret}, 759

\bibitem[{{Gierli{\'n}ski} \& {Done}(2006)}]{2006MNRAS.371L..16G}
{Gierli{\'n}ski}, M. \& {Done}, C. 2006, \mnras, 371, L16

\bibitem[{{Kalberla} {et~al.}(2005){Kalberla}, {Burton}, {Hartmann}, {Arnal},
  {Bajaja}, {Morras}, \& {P{\"o}ppel}}]{2005A&A...440..775K}
{Kalberla}, P.~M.~W., {Burton}, W.~B., {Hartmann}, D., {et~al.} 2005, \aap,
  440, 775

\bibitem[{{Kallman} \& {Bautista}(2001)}]{2001ApJS..133..221K}
{Kallman}, T. \& {Bautista}, M. 2001, \apjs, 133, 221

\bibitem[{{Krongold} {et~al.}(2007){Krongold}, {Nicastro}, {Elvis},
  {Brickhouse}, {Binette}, {Mathur}, \&
  {Jim{\'e}nez-Bail{\'o}n}}]{2007ApJ...659.1022K}
{Krongold}, Y., {Nicastro}, F., {Elvis}, M., {et~al.} 2007, \apj, 659, 1022

\bibitem[{{Lamastra} {et~al.}(2006){Lamastra}, {Perola}, \&
  {Matt}}]{2006A&A...449..551L}
{Lamastra}, A., {Perola}, G.~C., \& {Matt}, G. 2006, \aap, 449, 551

\bibitem[{{Lasota} {et~al.}(1996){Lasota}, {Abramowicz}, {Chen}, {Krolik},
  {Narayan}, \& {Yi}}]{1996ApJ...462..142L}
{Lasota}, J., {Abramowicz}, M.~A., {Chen}, X., {et~al.} 1996, \apj, 462, 142

\bibitem[{{Liu} \& {Taam}(2009)}]{2009ApJ...707..233L}
{Liu}, B.~F. \& {Taam}, R.~E. 2009, \apj, 707, 233

\bibitem[{{Longinotti} {et~al.}(2008){Longinotti}, {de La Calle}, {Bianchi},
  {Guainazzi}, \& {Dov{\v c}iak}}]{2008MmSAI..79..259L}
{Longinotti}, A.~L., {de La Calle}, I., {Bianchi}, S., {Guainazzi}, M., \&
  {Dov{\v c}iak}, M. 2008, Memorie della Societa Astronomica Italiana, 79, 259

\bibitem[{{Lu} \& {Wang}(2000)}]{2000ApJ...537L.103L}
{Lu}, Y. \& {Wang}, T. 2000, \apjl, 537, L103

\bibitem[{{Marconi} {et~al.}(2004){Marconi}, {Risaliti}, {Gilli}, {Hunt},
  {Maiolino}, \& {Salvati}}]{2004MNRAS.351..169M}
{Marconi}, A., {Risaliti}, G., {Gilli}, R., {et~al.} 2004, \mnras, 351, 169

\bibitem[{{Markowitz} \& {Reeves}(2009)}]{2009ApJ...705..496M}
{Markowitz}, A.~G. \& {Reeves}, J.~N. 2009, \apj, 705, 496

\bibitem[{{Matt} {et~al.}(2005){Matt}, {Porquet}, {Bianchi}, {Falocco},
  {Maiolino}, {Reeves}, \& {Zappacosta}}]{2005A&A...435..857M}
{Matt}, G., {Porquet}, D., {Bianchi}, S., {et~al.} 2005, \aap, 435, 857

\bibitem[{{Meyer} {et~al.}(2000){Meyer}, {Liu}, \&
  {Meyer-Hofmeister}}]{2000A&A...361..175M}
{Meyer}, F., {Liu}, B.~F., \& {Meyer-Hofmeister}, E. 2000, \aap, 361, 175

\bibitem[{{Narayan} \& {McClintock}(2008)}]{2008NewAR..51..733N}
{Narayan}, R. \& {McClintock}, J.~E. 2008, New Astronomy Review, 51, 733

\bibitem[{{Narayan} \& {Yi}(1994)}]{1994ApJ...428L..13N}
{Narayan}, R. \& {Yi}, I. 1994, \apjl, 428, L13

\bibitem[{{Nelson}(2000)}]{2000ApJ...544L..91N}
{Nelson}, C.~H. 2000, \apjl, 544, L91

\bibitem[{{Netzer} {et~al.}(2003){Netzer}, {Kaspi}, {Behar}, {Brandt},
  {Chelouche}, {George}, {Crenshaw}, {Gabel}, {Hamann}, {Kraemer}, {Kriss},
  {Nandra}, {Peterson}, {Shields}, \& {Turner}}]{2003ApJ...599..933N}
{Netzer}, H., {Kaspi}, S., {Behar}, E., {et~al.} 2003, \apj, 599, 933

\bibitem[{{Quataert} {et~al.}(1999){Quataert}, {Di Matteo}, {Narayan}, \&
  {Ho}}]{1999ApJ...525L..89Q}
{Quataert}, E., {Di Matteo}, T., {Narayan}, R., \& {Ho}, L.~C. 1999, \apjl,
  525, L89

\bibitem[{{Revnivtsev} {et~al.}(2004){Revnivtsev}, {Sazonov}, {Jahoda}, \&
  {Gilfanov}}]{2004A&A...418..927R}
{Revnivtsev}, M., {Sazonov}, S., {Jahoda}, K., \& {Gilfanov}, M. 2004, \aap,
  418, 927

\bibitem[{{Risaliti} {et~al.}(2002){Risaliti}, {Elvis}, \&
  {Nicastro}}]{2002ApJ...571..234R}
{Risaliti}, G., {Elvis}, M., \& {Nicastro}, F. 2002, \apj, 571, 234

\bibitem[{{Ross} \& {Fabian}(2005)}]{2005MNRAS.358..211R}
{Ross}, R.~R. \& {Fabian}, A.~C. 2005, \mnras, 358, 211

\bibitem[{{Sazonov} {et~al.}(2007){Sazonov}, {Revnivtsev}, {Krivonos},
  {Churazov}, \& {Sunyaev}}]{sazonov07}
{Sazonov}, S., {Revnivtsev}, M., {Krivonos}, R., {Churazov}, E., \& {Sunyaev},
  R. 2007, \aap, 462, 57

\bibitem[{{Str{\"u}der} {et~al.}(2001){Str{\"u}der}, {Briel}, {Dennerl},
  {Hartmann}, {Kendziorra}, {Meidinger}, {Pfeffermann}, {Reppin}, {Aschenbach},
  {Bornemann}, {Br{\"a}uninger}, {Burkert}, {Elender}, {Freyberg}, {Haberl},
  {Hartner}, {Heuschmann}, {Hippmann}, {Kastelic}, {Kemmer}, {Kettenring},
  {Kink}, {Krause}, {M{\"u}ller}, {Oppitz}, {Pietsch}, {Popp}, {Predehl},
  {Read}, {Stephan}, {St{\"o}tter}, {Tr{\"u}mper}, {Holl}, {Kemmer}, {Soltau},
  {St{\"o}tter}, {Weber}, {Weichert}, {von Zanthier}, {Carathanassis}, {Lutz},
  {Richter}, {Solc}, {B{\"o}ttcher}, {Kuster}, {Staubert}, {Abbey}, {Holland},
  {Turner}, {Balasini}, {Bignami}, {La Palombara}, {Villa}, {Buttler},
  {Gianini}, {Lain{\'e}}, {Lumb}, \& {Dhez}}]{pn}
{Str{\"u}der}, L., {Briel}, U., {Dennerl}, K., {et~al.} 2001, \aap, 365, L18

\bibitem[{{Stuhlinger} {et~al.}(2006){Stuhlinger}, {Altieri}, {Esquej},
  {Kirsch}, {Metcalfe}, {Pollock}, {Saxton}, {Smith}, {Talavera}, {Sembay},
  {Read}, {Baskill}, {Haberl}, {Freyberg}, {Dennerl}, {Kaastra}, {den Herder},
  {de Vries}, \& {Vink}}]{2006ESASP.604..937S}
{Stuhlinger}, M., {Altieri}, B., {Esquej}, M.~P., {et~al.} 2006, in ESA Special
  Publication, Vol. 604, The X-ray Universe 2005, ed. {A.~Wilson}, 937--+

\bibitem[{{Svoboda} {et~al.}(2009){Svoboda}, {Dov{\v c}iak}, {Goosmann}, \&
  {Karas}}]{2009A&A...507....1S}
{Svoboda}, J., {Dov{\v c}iak}, M., {Goosmann}, R., \& {Karas}, V. 2009, \aap,
  507, 1

\bibitem[{{Takata} {et~al.}(1994){Takata}, {Yamada}, {Saito}, {Chamaraux}, \&
  {Kazes}}]{takata94}
{Takata}, T., {Yamada}, T., {Saito}, M., {Chamaraux}, P., \& {Kazes}, I. 1994,
  \aaps, 104, 529

\bibitem[{{Tueller} {et~al.}(2008){Tueller}, {Mushotzky}, {Barthelmy},
  {Cannizzo}, {Gehrels}, {Markwardt}, {Skinner}, \&
  {Winter}}]{2008ApJ...681..113T}
{Tueller}, J., {Mushotzky}, R.~F., {Barthelmy}, S., {et~al.} 2008, \apj, 681,
  113

\bibitem[{{Turner} {et~al.}(2001){Turner}, {Abbey}, {Arnaud}, {Balasini},
  {Barbera}, {Belsole}, {Bennie}, {Bernard}, {Bignami}, {Boer}, {Briel},
  {Butler}, {Cara}, {Chabaud}, {Cole}, {Collura}, {Conte}, {Cros}, {Denby},
  {Dhez}, {Di Coco}, {Dowson}, {Ferrando}, {Ghizzardi}, {Gianotti}, {Goodall},
  {Gretton}, {Griffiths}, {Hainaut}, {Hochedez}, {Holland}, {Jourdain},
  {Kendziorra}, {Lagostina}, {Laine}, {La Palombara}, {Lortholary}, {Lumb},
  {Marty}, {Molendi}, {Pigot}, {Poindron}, {Pounds}, {Reeves}, {Reppin},
  {Rothenflug}, {Salvetat}, {Sauvageot}, {Schmitt}, {Sembay}, {Short},
  {Spragg}, {Stephen}, {Str{\"u}der}, {Tiengo}, {Trifoglio}, {Tr{\"u}mper},
  {Vercellone}, {Vigroux}, {Villa}, {Ward}, {Whitehead}, \& {Zonca}}]{mos}
{Turner}, M.~J.~L., {Abbey}, A., {Arnaud}, M., {et~al.} 2001, \aap, 365, L27

\bibitem[{{Turner} {et~al.}(1997){Turner}, {George}, {Nandra}, \&
  {Mushotzky}}]{1997ApJS..113...23T}
{Turner}, T.~J., {George}, I.~M., {Nandra}, K., \& {Mushotzky}, R.~F. 1997,
  \apjs, 113, 23

\bibitem[{{Urry} \& {Padovani}(1995)}]{1995PASP..107..803U}
{Urry}, C.~M. \& {Padovani}, P. 1995, \pasp, 107, 803

\bibitem[{{Wang} {et~al.}(2006){Wang}, {Wei}, \& {He}}]{2006ApJ...638..106W}
{Wang}, J., {Wei}, J.~Y., \& {He}, X.~T. 2006, \apj, 638, 106

\bibitem[{{Wilms} {et~al.}(2000){Wilms}, {Allen}, \& {McCray}}]{wilms00}
{Wilms}, J., {Allen}, A., \& {McCray}, R. 2000, \apj, 542, 914

\bibitem[{{Yuan} \& {Narayan}(2004)}]{2004ApJ...612..724Y}
{Yuan}, F. \& {Narayan}, R. 2004, \apj, 612, 724

\end{thebibliography}

\end{document}